 \documentclass[final,5p,times,twocolumn]{elsarticle}

\usepackage{amssymb}
\usepackage{hyperref,amssymb,amsmath,mathrsfs,bm,graphicx}
\usepackage[dvipsnames]{xcolor}
\usepackage{comment}

\usepackage[dvipsnames]{xcolor}

\def\be{\begin{equation}}
\def\ee{\end{equation}}
\def\bea{\begin{eqnarray}}
\def\eea{\end{eqnarray}}

\def\yzero{\smash{\hbox{$y\kern-4pt\raise1pt\hbox{${}^\circ$}$}}}

\def\beq{\begin{equation}}
\def\eeq{\end{equation}}
\def\beqa{\begin{eqnarray}}
\def\eeqa{\end{eqnarray}}

\def\-{\hphantom{-}}

\def\s2{\frac{1}{\sqrt2}}

\def\beq{\begin{equation}}
\def\eeq{\end{equation}}
\def\beqa{\begin{eqnarray}}
\def\eeqa{\end{eqnarray}}

\def\IF{\relax{\rm I\kern-.18em F}}
\def\II{\relax{\rm I\kern-.18em I}}
\def\IP{\relax{\rm I\kern-.18em P}}
\def\IC{\relax\hbox{\kern.25em$\inbar\kern-.3em{\rm C}$}}
\def\IR{\relax{\rm I\kern-.18em R}}

\def\Dsl{\,\raise.15ex\hbox{/}\mkern-13.5mu D} %this one can be subscripted
\def\IZ{Z\kern-.4em  Z}

\journal{Physics Letters B}

\begin{document}

\begin{frontmatter}



\title{Supersymmetric algebra of the massive supermembrane}


\author[label1,label2]{M.P. Garcia del Moral}
\ead{m-pilar.garciam@unirioja.es; maria.garciadelmoral@uantof.cl}
\author[label2]{P. León \fnref{1}}
\ead{pablo.leon@ua.cl}
\author[label2]{A. Restuccia}
\ead{alvaro.restuccia@uantof.cl}

\affiliation[label1]{organization={Área de Física, Departamento de Química, Universidad de la Rioja},%Department and Organization
            addressline={ La Rioja 26006}, 
            country={Spain}}          

\affiliation[label2]{organization={Departamento de Física, Universidad de Antofagasta},%Department and Organization
           addressline={Aptdo 02800}, 
           country={Chile}}            
\fntext[1]{All authors have contributed equally to this work}
\begin{abstract}
In this paper, we obtain the explicit expression of the supersymmetric algebra associated with the recently proposed massive supermembrane including all surface terms. We formulate the theory as the limit of a supermembrane on a genus-two compact Riemann surface when one of the handles becomes a string attached to a torus. The formulation reduces to a supermembrane on a punctured torus with a "string spike" (in the sense of  \cite{dwln}), attached to it. In this limit, we identify all surface terms of the algebra and give the explicit expression of the Hamiltonian in agreement with the previous formulation of it. The symmetry under area preserving diffeomorphisms,  connected and nonconnected to the identity, is also discussed. Only parabolic $Sl(2,\mathbb{Z})$ discrete symmetries are preserved.
\end{abstract}




\end{frontmatter}

\section{Introduction}
Recently, new aspects of M2-brane theory in D=11 dimensions have been developed. In \cite{Nicolai2} using the Nicolai map, a perturbative quantization approach has been proposed. In \cite{mpgm15} the existence and uniqueness of the ground state of the theory on the valleys of the theory have been obtained, In \cite{mpgm,Dasgupta,mpgm6,mpgm14} new sectors of the theory formulated on different backgrounds characterized by topological conditions have been analyzed. In contrast to the formulations on a Minkowski target space, these supersymmetric sectors of the M2-brane have a discrete spectrum. They correspond to supermembranes with a topological condition associated with the presence of 2-form worldvolume fluxes induced by either the presence of a topological central charge condition \cite{mpgm}, the presence of supergravity constant and quantized three-form \cite{mpgm6}, or either on a target space with $G_4$ content, as the supermembrane on a pp-wave \cite{Dasgupta} whose matrix model corresponds to \cite{Maldacena3}, or more recently the formulation of a massive supermembrane \cite{mpgm14}.This massive supermembrane corresponds to a supermembrane theory formulated on a $M_9\times LCD$ background, where $M_9$ is a nine dimensional Minkowski space and $LCD$ is a Light Cone Diagram, a two dimensional flat strip with identifications and with prescribed segments whose curvature becomes infinite at some points. This surface only has one (non-trivial) compact dimension, and therefore the supermembrane in this background exhibits ten non-compact dimensions. Moreover, the theory has nontrivial mass terms not present in the supermembrane theory compactified on a circle that, together with the rest of the structure of the potential, render the spectrum of the regularized theory to be discrete. The goal of this paper is to characterize the susy algebra including all the boundary terms. This may shed light on the role of the singularities in the structure of the constraints that will be useful to obtain the string theory associated with this sector. Specifically, from the closure of the algebra, one can infer all the global constraints of the supermembrane. In the approach we follow in this paper, the constraints arise directly from the geometrical construction in agreement with the consequences of the closure of the algebra. Also, in general, the algebra will determine the symmetries of the supermembrane theory and the inherited symmetries of the associated string theory. Moreover, after a double-dimensional reduction, one of these constraints yields to the closed string level matching condition (see \cite{duff,lc}). Furthermore, the analysis of the singularities allows to characterize the dimensions of the sources coupled to the M2 -brane, as for the example the M9-brane discussed in \cite{Bergshoeff7}. The supermembrane only admits backgrounds that allow a consistent coupling to the 11D supergravity and its reductions. Hence, the algebra may give light to the supergravity background to which this massive supermembrane couples.This can be obtained by analysing the zero mode structure of the algebra \cite{dwhn}. 
The study of algebras and their deformations, and consequently their symmetries, have also been used in the literature to obtain kinetic terms of their associated supergravity Lagrangian densities in the context of limits of GR gravity, see for example \cite{Bergshoeff11, Gomis20, Bergshoeff17} as well as in the context of supergravity ones, see \cite{M-algebra,Edelstein06, Ravera22}. Although we will not proceed in this direction, this is another possible application of the results of this work.

The paper is structured as follows: In section 2, we recall the basic aspects of the supermembrane theory formulation and its Hamiltonian in the case of a supermembrane with a topological central charge condition. In section 3, we summarize the main properties of the Light Cone Diagram formulation that will be needed for the computations. In section 3, we present a new formulation of the massive supermembrane obtained in \cite{mpgm14} which directly incorporates all the boundary terms of the formulation. In section 5, we obtain the supersymmetric transformation, and in section 6, we get the supersymmetric algebra of supercharges. In section 7, we discuss another fundamental symmetry , that is, the area preserving diffeomorphisms, in order to characterize completely the symmetries of the theory. In section 8, we present our conclusions.
%%%%%%%%%%%%%%%%%%%%%%%%%%%%%%%%%%%%%%%%%%%%%%%%%%%%%%%
\section{The supermembrane action in the light cone gauge}

The supermembrane was originally introduced in \cite{Bergshoeff}. Its formulation in the Light Cone Gauge (LCG) on a Minkowski target space was obtained \cite{dwhn}.
In this section we will briefly review some of those results in \cite{dwhn} and we will present the supermembrane action  in the light cone gauge on $M_9 \times T^2$. The action of the supermembrane in a Minkowski space-time is given by 

\begin{eqnarray}
S &=& -T_{M2} \int_{R\times \Sigma} d\xi^3 \bigg[\sqrt{-g} +\varepsilon^{uvw} \bar{\tilde{\Psi}} \Gamma_{\mu \nu} \partial_{w}\tilde{\Psi}  \nonumber \\ &\times&\bigg(\frac{1}{2}\partial_u \tilde{X}^\mu (\partial_v \tilde{X}^\nu + \bar{\tilde{\Psi}}\Gamma^\nu \partial_v \tilde{\Psi}) + \frac{1}{6} \bar{\tilde{\Psi}}\Gamma^\mu \partial_u \tilde{\Psi} \bar{\tilde{\Psi}}\Gamma^\nu\partial_v\tilde{\Psi}\bigg)\bigg], \nonumber \\ &&
\end{eqnarray}
where $T_{M2}$ is the M2-brane tension, $\Gamma^\mu$ are the gamma matrix in eleven dimensions, $\tilde{X}^\mu$ $(\mu,\nu = 0,..,10)$ are the embedding maps of the supermembrane, $\theta$ is a 32 component Majorana spinor and $\Sigma$ is a compact Riemann surface. All the fields are functions of the world-volume coordinates $\xi^u$ $(u,v,w = 0,1,2)$ and $g_{uv}$ are the components of the world-volume induced metric, this is

\begin{eqnarray}
g_{uv} = (\partial_u \tilde{X}^\mu + \bar{\tilde{\Psi}}\Gamma^\mu \partial_u \tilde{\Psi})(\partial_v \tilde{X}^\nu + \bar{\tilde{\Psi}}\Gamma^\nu \partial_v \tilde{\Psi}) \eta_{\mu \nu}.
\end{eqnarray}

Now we can use the light cone coordinates $\tilde{X}^\mu = (X^+,X^-,\tilde{X}^{M})$ with $M,N = 1,..,9$

\begin{eqnarray}
X^{\pm} = \frac{1}{\sqrt{2}}(\tilde{X}^{10} \pm \tilde{X}^0), \quad \Gamma^{\pm} = \frac{1}{\sqrt{2}}(\Gamma^{10} \pm \Gamma^0),
\end{eqnarray}
and, decomposing $\xi^u=(t,\sigma^r)$ with $r=1,2$, one can fix the LCG as follows,

\begin{eqnarray}
X^+ = t, \quad \Gamma^+ \tilde{\Psi} = 0. 
\end{eqnarray}
Thus, the Lagrangian density can be written as \footnote{We are using $\varepsilon^{0rs} = -\epsilon^{rs}$}

\begin{eqnarray}
\mathcal{L} &=& -T_{M2} (\sqrt{\bar{g}\Delta} + \epsilon^{rs}\partial_r \tilde{X}^{M} \bar{\tilde{\Psi}}\Gamma^-\Gamma_M\partial_s \tilde{\Psi}),
\end{eqnarray}
where 

\begin{eqnarray*}
\bar{g}_{rs} &=& \partial_r \tilde{X}^{M} \partial_s \tilde{X}_M, \nonumber \\  
u_r = g_{0r} &=& \partial_r \tilde{X}^- + \partial_t \tilde{X}^{M}\partial_r \tilde{X}_M + \tilde{\Psi}\Gamma^-\partial_r \tilde{\Psi}, \\
g_{00} &=& 2\partial_tX^-+\partial_t \tilde{X}^{M}\partial_t\tilde{X}_M + 2\bar{\tilde{\Psi}}\Gamma^-\partial_0\tilde{\Psi},
\end{eqnarray*}
and $\bar{g} = det(\bar{g}_{rs})$, $\Delta = -g_{00} + u_r \bar{g}^{rs}u_s$. Then, the conjugate momenta can be written as

\begin{align*}
\small
& \tilde{P}_- = T_{M2} \sqrt{\frac{\bar{g}}{\Delta}}, \quad \tilde{P}^M = \tilde{P}_- (\partial_0\tilde{X}^{M}-u_rg^{rs}\partial_s\tilde{X}^{M}), \quad \tilde{S} = - \tilde{P}_-\Gamma^-\tilde{\Psi}.
\end{align*}
Thus, the Hamiltonian density is given by

\begin{eqnarray}
\mathcal{H} = \frac{\tilde{\textbf{P}}^2+T^2_{M2}\bar{g}}{2\tilde{P}_-} - T_{M2}\epsilon^{rs}\partial_r\tilde{X}^{M} \bar{\tilde{\Psi}} \Gamma^-\Gamma_M\partial_s \tilde{\Psi},
\end{eqnarray}
subject to the following primary constraints 
\begin{align}\label{fc0}
 & \Phi_r = \tilde{\textbf{P}} \partial_r \tilde{\textbf{X}} + \tilde{P}_-\partial_r \tilde{X}^- + \bar{\tilde{S}}\Gamma^-\tilde{\Psi} = 0, \\ 
 & \Upsilon = \tilde{S} + T_{M2}\sqrt{\frac{\bar{g}}{\Delta}}\Gamma^-\tilde{\Psi} = 0.
\end{align}
The Dirac analysis of these constraints yields that $\Phi_r$ is of first class while $\Upsilon$ is of second class.

Now, we can use the area preserving diffeomorphims to set the gauge $\tilde{P}_- = P_-^0\sqrt{W}$, where $\sqrt{\tilde{W}}$ is a scalar density satisfying 

\begin{eqnarray}
\int_\Sigma \sqrt{\tilde{W}} = 1.
\end{eqnarray}
This allows us to introduce the Lie bracket

\begin{eqnarray}
\{\cdot, \cdot \} = \frac{\epsilon^{rs}}{\sqrt{\tilde{W}}}\partial_r \cdot \partial_s \cdot.
\end{eqnarray}

The supermembrane Lagrangian density can be written in a way that is explicitly invariant under area preserving diffeomorphism (see \cite{dwhn}). This requires the introduction of a gauge field $\omega$ related to time-dependent reparametrizations of the world-volume. This is

\begin{eqnarray}\label{SL}
   \frac{\mathcal{L}}{P_0^+ \sqrt{\tilde{W}}} &=& \frac{1}{2}(D_0\tilde{X}^{M})^2+\bar{\tilde{\Psi}}\Gamma^- D_0\tilde{\Psi} - \frac{T_{M2}^2}{4P_0^+}\{\tilde{X}^{M}, \tilde{X}^N\}^2  \nonumber \\ &+& \frac{T_{M2}}{P_0^+}\bar{\tilde{\Psi}}\Gamma^-\Gamma^a \{\tilde{X}^{M},\tilde{\Psi}\} +D_0 \tilde{X}^-,
\end{eqnarray}
where

\begin{eqnarray}
   D_0 \bullet = \partial_t \bullet - \{\omega, \bullet\}, \quad \{\bullet,\bullet\} = \frac{\epsilon^{rs}}{\sqrt{W}}\partial_r \bullet \partial_s \bullet.
\end{eqnarray}
Furthermore, we can now solve (\ref{fc0}) for $\tilde{X}^-$, this is 

\begin{eqnarray}\label{xmso}
\partial_r \tilde{X}^- = -\frac{1}{P_-^0\sqrt{\tilde{W}}} (\tilde{\textbf{P}} \partial_r \tilde{\textbf{X}}+\bar{\tilde{S}}\Gamma^-\partial_r\tilde{\Psi}).
\end{eqnarray}
The integrability conditions for the existence of a single valued $\tilde{X}^-$ solution of (\ref{xmso}) are: first, since the left-hand side of (\ref{xmso}) can be expressed as a closed form, then the same must happen for the right-hand side member. This condition yields a local constraint for the right-hand member that must be satisfied at each point of $\Sigma$,i.e.

\begin{eqnarray}
 \phi &=& d(d\tilde{X}^-) = d\left[\frac{1}{\sqrt{\tilde{W}}} (\tilde{\textbf{P}} d \tilde{\textbf{X}}+\bar{\tilde{S}}\Gamma^-d\tilde{\Psi})\right] = 0. 
\end{eqnarray}
The second condition, since the left-hand member is an exact form ($\tilde{X}^-$ is single valued), then the right-hand member must also be exact. This restriction is imposed by taking the periods around the homology basis to be zero, i.e.
\begin{eqnarray}
 \varphi_k &=& \int_{\mathcal{C}_k} d\tilde{X}^- = \int_{\mathcal{C}_k}\frac{1}{\sqrt{\tilde{W}}} (\tilde{\textbf{P}} d \tilde{\textbf{X}}+\bar{\tilde{S}}\Gamma^-d\tilde{\Psi}) =0,    
\end{eqnarray}
where $\mathcal{C}_k$ ($k = 1,..,2g$ for $g>1$)are the homology basis of one-cycles over $\Sigma$. 
They correspond to the local and global first class constraints associated with the residual symmetry of Area Preserving Diffeomorphisms (APD).

Now it is possible to write the Hamiltonian of the theory as,
\begin{eqnarray} \label{g2H}
H &=& \frac{1}{2P_-^0} \int_\Sigma d^2\sigma \sqrt{\tilde{W}} \bigg[\left(\frac{\tilde{\textbf{P}}}{\sqrt{\tilde{W}}}\right)^2 + \frac{T^2_{M2}}{2}\{\tilde{X}^{M},\tilde{X}^N\}^2 \nonumber \\  &-& 2T_{M2} P_-^0 \bar{\tilde{\Psi}}\Gamma^-\Gamma_M \{\tilde{X}^{M},\tilde{\Psi}\}\bigg].
\end{eqnarray}

Now one can compactify the M2-brane Hamiltonian on $M_9 \times T^2$ and take as a base manifold a regular genus-two Riemann surface $\Sigma_2$. Thus, due to the compact dimensions, the embedding maps can be decomposed as $\tilde{X}^{M} = (\tilde{X}^m,\tilde{X}^r)$, with $m=1,...,7$ labeling the noncompact dimensions and $r=1,2$ the compact ones associated with the 2-torus. The $\tilde{X}^m$ maps $\Sigma_2$ to the transverse subspace of $M_9$ while $\tilde{X}^r$ maps $\Sigma_2$ to the target $T2$.

 Hence, the Hamiltonian of the supermembrane can be written as

\begin{align}
   &H = \frac{1}{2P_-^0} \int_{\Sigma_2} d^2\sigma \sqrt{\tilde{W}} \bigg[\left(\frac{\tilde{P}_m}{\sqrt{\tilde{W}}}\right)^2 + \left(\frac{\tilde{P}_r}{\sqrt{\tilde{W}}}\right)^2 + \frac{T^2_{M2}}{2}\{\tilde{X}^m,\tilde{X}^n\}^2 \nonumber \\ &  + T^2_{M2}\{\tilde{X}^m,\tilde{X}^r\}^2  + \frac{T^2_{M2}}{2}\{\tilde{X}^r,\tilde{X}^s\}^2 -2T_{M2} P_-^0 \bar{\tilde{\Psi}}\Gamma^-\Gamma_m \{\tilde{X}^m,\tilde{\Psi}\} \nonumber \\ & - 2T_{M2} P_-^0 \bar{\tilde{\Psi}}\Gamma^-\Gamma_r \{\tilde{X}^r,\tilde{\Psi}\}\bigg], 
\end{align}
 subject now to the following  ive APD constraints

\begin{eqnarray}
 \phi &=& d\left[\frac{1}{\sqrt{\tilde{W}}} (\tilde{P}_m d\tilde{X}^m +\tilde{P}_r d\tilde{X}^r + \bar{\tilde{S}}\Gamma^-d\tilde{\Psi})\right] = 0, \\ 
 \varphi_k &=& \int_{\mathcal{C}_k}\frac{1}{\sqrt{\tilde{W}}} (\tilde{P}_m d\tilde{X}^m +\tilde{P}_r d\tilde{X}^r +\bar{\tilde{S}}\Gamma^-d\tilde{\Psi}) =0, 
\end{eqnarray}
where $k=1,...,4$.

We will use these expressions in the subsequent sections of the paper. 
%%%%%%%%%%%%%%%%%%%%%%%%%%%%%%%%%%%%%%%%%%%%%%%%%%%%%%%%%%%%%%%%%%
\section{Parametrization of the twice punctured torus}

\begin{figure*}
    \centering
    \includegraphics[scale=0.4]{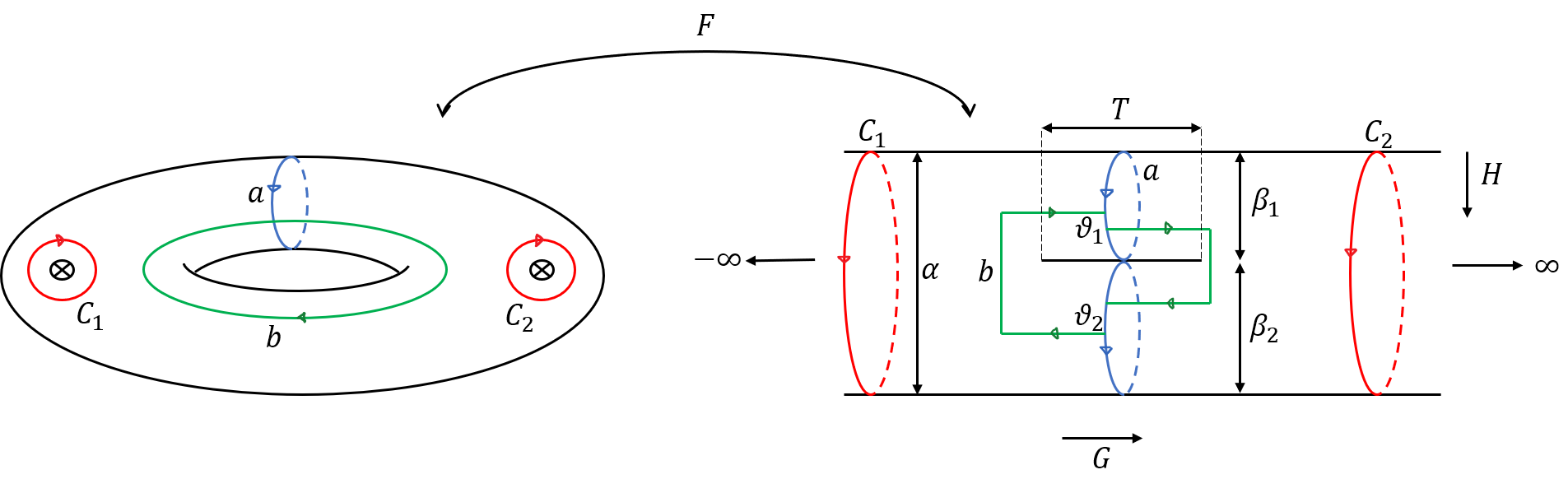}
    \caption{The torus with two punctures and the one loop interaction string diagram with one incoming/outgoing string. The Mandelstam map send the punctures over the torus to $\pm\infty$ in the LCD.}
    \label{fig:equi}
\end{figure*}

In this section, we recall some useful results about the relation between the Light Cone diagram (LCD) and the torus with two punctures $\Sigma_{1,2}$ (see figure (\ref{fig:equi}))  needed to describe the massive supermembrane formulation. The Light Cone diagram is a two dimensional flat strip with identifications and 
prescribed segments whose curvature becomes infinite at some points. These results are the base of the massive supermembrane formulation \cite{mpgm14} and they will be useful in the next sections. The relation between these two surfaces is given by the Mandelstam map (see \cite{Mandelstam,Giddings2})

\begin{eqnarray}\label{Mm}
    F(z)&=&\alpha \ln\left[\frac{\Theta_1(z-Z_1\vert \tau)}{\Theta_1(z-Z_2\vert \tau)}\right]-2\pi i \alpha\frac{Im (Z_1-Z_2)}{Im \tau}(z-z_0), \nonumber \\ &&
\end{eqnarray}
where $\Theta_1(z,\tau)$ are the Jacobi functions and $Z_r$ with $r=1,2$ are the positions of the punctures in a complex coordinates over the torus. The set of parameters necessary to characterize the torus with two punctures are the Teichm\"uller parameter, $\tau$, and the positions of the Punctures, $Z_r$. On the twice punctured torus the coordinate system $z$ is defined in terms of the holomorphic one-form $dz$ satisfying

\begin{eqnarray}
dz = d\hat{X}^1 + \tau d\hat{X}^2, \quad \mbox{with} \quad \int_{\mathcal{C}_k} d\hat{X}^r = \delta_k^r,
\end{eqnarray}
where $d\hat{X}^r$ is a set of real normalized forms over the regular torus.

On the other hand, the set of parameters that describe the LCD is given by the external momenta $\alpha$, the internal momenta $\beta_r$, the interaction time $T$, and the twist angles $\theta_r$. Then in order to complete the equivalence between the two surfaces, (see figure 1.),  the following relation between both sets of parameters is required 

\begin{equation}
2\pi i (Z_1-Z_2)= (\theta_1+\theta_2)\beta_1-\alpha\theta_2-2\pi i\beta_1 \tau.
\end{equation}  
It is useful to decompose the Mandelstam map in terms of its real and imaginary parts, that is $F= G+iH$. The function $G$ is single valued, but $dG$ is harmonic, since it has poles at the punctures. The function $H$ is multivalued and $dH$ is harmonic.The behavior of each function near the punctures is given by
\begin{eqnarray}
G  &\sim&  (-1)^{r+1}\alpha\ln\vert z-Z_r\vert, \\
H &\sim& (-1)^{r+1}\alpha \varphi, \quad \mbox{with} \quad \varphi\in (0,2\pi) \ \ (r=1,2). \label{HB}
\end{eqnarray}
On the other hand, near the zeros of $dF$,denoted as $P_a$, the functions $G$ and $H$ can be written as
\begin{eqnarray}
G(z)-G(P_a) &\sim &  \frac{1}{2}Re(D(P_a)(z-P_a)^2),\label{Gz}  \\
H(z)-H(P_a)  & \sim &  \frac{1}{2}Im(D(P_a)(z-P_a)^2), \label{Hz}
\end{eqnarray}
where
\begin{align*}
&D(P_a) = \sum_{r=1}^{2} (-1)^{r+1} \bigg[ \frac{\partial^2_z \Theta_1(P_a-z_r,\tau)}{\Theta_1(P_a-z_r,\tau)} -\left(\frac{\partial_z \Theta_1(P_a-z_r,\tau)}{\Theta_1(P_a-z_r,\tau)}\right)^2 \bigg].
\end{align*}
Finally, we recall some  properties of the functions $K$ and $H$ that will be useful in the next section,

\begin{eqnarray}
    G(z+1)-G(z) &=& G(z+\tau)-G(z)=0  \label{Ks}\\
    H(z+1)-H(z) &=& 2\pi \alpha \frac{Im(Z_2-Z_1)}{Im(\tau)}, \label{Hs1} \\
    H(z+\tau)-H(z) & = & \frac{2\pi \alpha Im((Z_2-Z_1)\bar{\tau})}{Im(\tau)}. \label{Hs2}
\end{eqnarray}
%%%%%%%%%%%%%%%%%%%%%%%%%%%%%%%%%%%%%%%%%%%%%%%%%%%%%%%%%%%%%%%

\section{Massive supermembrane}
In this section, we present a new formulation of the massive supermembrane and its connection with the formulation found in \cite{mpgm14}. Specifically, in order to make clearer the surface terms that appear in the supersymmetric algebra, we use a different approach than  \cite{mpgm14}. Instead of considering the supermembrane formulated in $M_9\times LCD$ on a twice punctured torus as the base manifold, we will start with the M2-brane on a compact genus-two Riemann surface $\Sigma_2$ as the base manifold in $M_9\times T^2$ as the target space. 
\begin{figure*}
    \centering
    \includegraphics[scale=0.3]{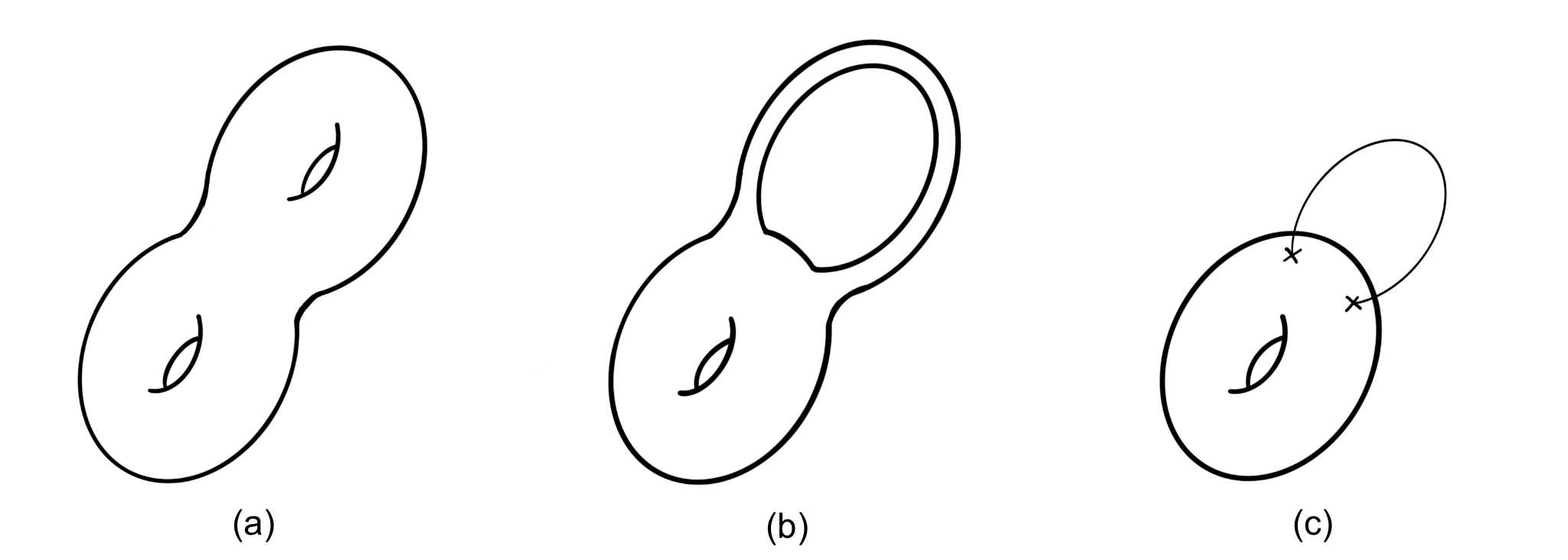}
    \caption{(a) The genus two regular Riemann surface $\Sigma_2$. (b) A deformation of $\Sigma_2$. (c) The surface $\tilde{\Sigma}_{1,2}$ obtained by taking one of the radii of $\Sigma_2$ tending to zero. This correspond to a singular $T^2$  with a string attached to it.}
    \label{fig:TD}
\end{figure*}
In order to establish a connection with the formulation of the massive supermembrane \cite{mpgm14}, we will take a specific limit to deform $\Sigma_2$ as described in the figure (\ref{fig:TD}). That is, we will assume that one of the radii of the handles of the genus two surface tends to zero. As a result, we can expand the maps $X^m$, $X^r$, and $\Psi$ in a Fourier series and keep only the order zero of the variable associated with the small radius. Thus, under these considerations, the supermembrane maps will depend only on the coordinate along the handle (see figure (\ref{fig:TD})-(b)). In this way, we get a string-like configuration like the ones described in \cite{Nicolai}.  Thus, we will end up with a surface, that we will denote $\tilde{\Sigma}_{1,2}$, which is a twice punctured torus $\Sigma_{1,2}$ with a string attached to the punctures (see figure (\ref{fig:TD})-(c)). Then we will also deform the target $T^2$ to a $LCD$ surface. Thus, the metric that we shall define over the $LCD$ on the target is given by

\begin{eqnarray}
     ds^2 &=& l^2d\hat{G}^2+ dH^2 = dK^2 + \alpha^2 d\hat{H}^2 \label{mono0},
\end{eqnarray}
where $\hat{H} = H/\alpha, \quad \hat{G} = G/\alpha$ and $l$ is constant with length units.

Now we can describe the dependence of the M2-brane fields in two regions. The first one is the definition of the maps on $\Sigma_{1,2}$ and the second one is the string attached to it that we shall denote as $\gamma_2$. Then, given a coordinate system, $z$ (given in the previous section), over $\Sigma_{1,2}$ and defining as $u$ the coordinate associated to $\gamma_2$ we can write

\begin{eqnarray}
    (\tilde{X}^m,\tilde{\Psi}) = \left\lbrace 
    \begin{array}{ll}
        (X^m(t,z,\bar{z}),\Psi(t,z,\bar{z})) & \mbox{over} \quad \Sigma_{1,2}  \\
        (Y^m(t,u),\Theta(t,u)) &  \mbox{over} \quad \gamma_2
    \end{array}
    \right. , 
\end{eqnarray}
and
\begin{eqnarray}
    \tilde{X}^r = \left\lbrace 
    \begin{array}{ll}
        X^K(t,z,\bar{z})\delta^r_1 +  X^H(t,z,\bar{z})\delta^r_2& \mbox{over} \quad \Sigma_{1,2}  \\
        Y^r(t,u) &  \mbox{over} \quad \gamma_2
    \end{array}
    \right. .
\end{eqnarray}
The maps $X^K$ and $X^H$ are defined as in \cite{mpgm14}, i.e, 

\begin{eqnarray}
    X^K = K+A^K, \quad X^H = H + A^H,
\end{eqnarray}
where $m$ is an integer and the 1-forms $dA^K$, $dA^H$ are exact over $\Sigma_{1,2}$

Under all this consideration, as discussed in \cite{Nicolai,dwln}, the string we are considering does not change the supermembrane energy and therefore we can write
\begin{eqnarray}
H = \int_{\Sigma_2 \rightarrow \tilde{\Sigma}_{1,2}}\mathcal{H} = \int_{\Sigma_{1,2}}\mathcal{H}.
\end{eqnarray}

The string-like configuration that we are considering here has no M2-brane dynamics associated with it. This is so because it does not have any contribution to the Hamiltonian of the theory. Thus, without losing generality, we can impose 

\begin{align}\label{cts}
&Y^m_s(u,t) = const, \quad Y^r_s(u,t) = const, \quad \Theta_s(u,t)=const,
\end{align}
which implies 

\begin{eqnarray}
    X^m\bigg|_{Z_1}^{Z_2} = \Psi \bigg|_{Z_1}^{Z_2} = 0.
\end{eqnarray}
On the other hand, since the $Y^r_s(u,t)$ are single value functions, it is reasonable to consider that $A^K$ and $A^H$ are continuous functions of $Y^r$. Consequently, 

\begin{eqnarray}
    A^K\bigg|_{Z_1}^{Z_2} = A^H \bigg|_{Z_1}^{Z_2} = 0.
\end{eqnarray}

\begin{figure*}
    \centering
    \includegraphics[scale=0.4]{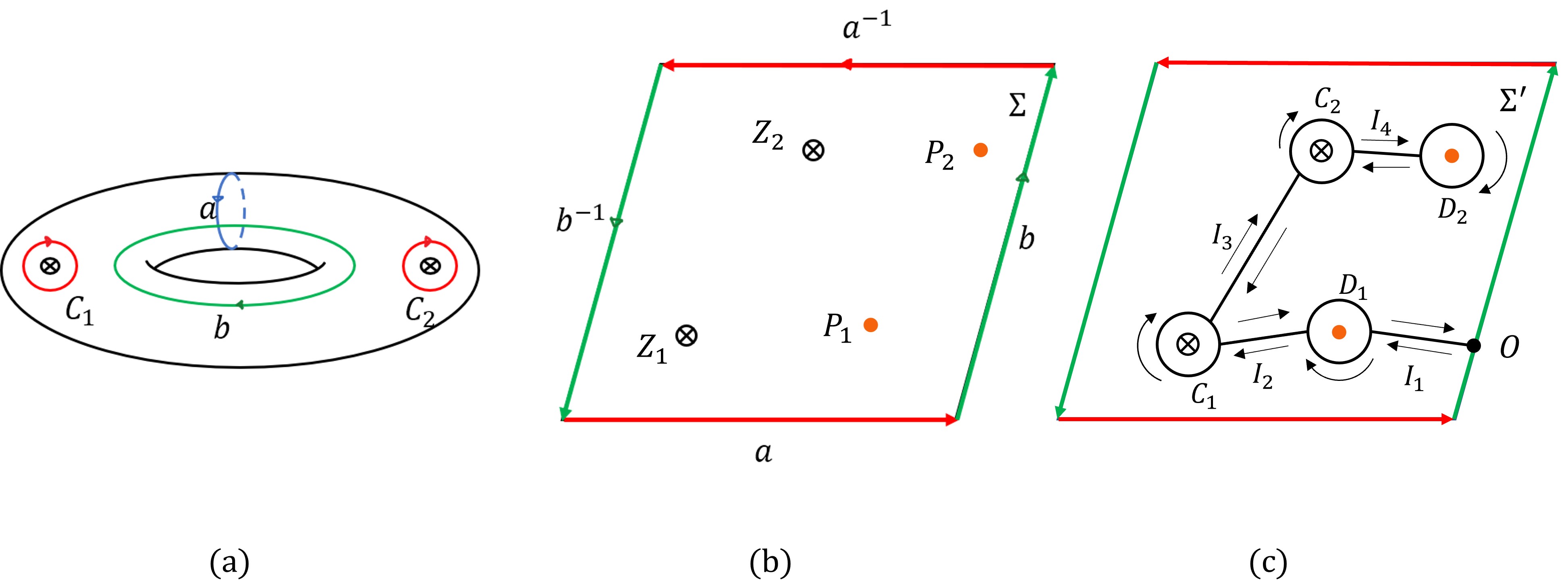}
    \caption{The region $\Sigma'$ obtained by cutting $\mathbf{\Sigma_{1,2}}$ through the curves $C_1$,$C_2$ and $I$. The path obtained by the union of the curves $C_1$,$I$,$C_2$ and $I^{-1}$ is denoted by $c$}
    \label{fig:SigmaP}
\end{figure*}

At this point, we can follow the same steps presented in \cite{mpgm14} to analyze the Hamiltonian over $\Sigma_{1,2}$. Specifically, we shall define the world-volume metric, over $\Sigma_{1,2}$, as 

\begin{eqnarray}
     \sqrt{W} &=& \frac{1}{4\pi}\epsilon^{rs}\partial_r \hat{K} \partial_s \hat{H},
\end{eqnarray}
where $K \equiv \tanh{\hat{G}}$. Then we can fix the gauge 

\bea
\{K,A^K\}+ m\{H,A^H\}=0. \label{gc}
\eea 
In order to deal with the singular behavior of the metric at the punctures and zeros we shall  cut the fundamental region of $\Sigma_{1,2}$, that we will call $\mathbf{\Sigma}_{1,2}$, through a closed curve that circumvents the two punctures, and the zeros with a radius $\epsilon$ and touch a point $O\in \partial \mathbf{\Sigma}_{1,2}$, see figure \ref{fig:SigmaP} (see \cite{farkas}). We shall denote as $C_r$ the curves around the punctures, $D_r$ the curves around the zeros, and as $I_j$, with $j=1,..,4$, to all the curves in between. Following the discussion presented in \cite{mpgm14}, it is clear that the curves $I_j$ can be chosen as curves $H=cte$. The we will denote as $\Sigma'$ the resulting region after cutting $\Sigma_{1,2}$. 

Under all these considerations, the Hamiltonian of the theory can be written as (see \cite{mpgm14} for more details)

\begin{align}\label{MM}
&H = \frac{(l  \alpha T_{M2} m )^2}{2P_0^+} + \frac{1}{2P_0^+}\lim_{\epsilon\rightarrow 0} \int_{\Sigma'}d\sigma^2\sqrt{W}\left[\left(\frac{P_m}{\sqrt{W}}\right)^2+\left(\frac{P_K}{\sqrt{W}}\right)^2 \right. \nonumber \\
    &  + \left(\frac{P_H}{\sqrt{W}}\right)^2 + T_{M2}^2\bigg(\frac{1}{2}\{X^m,X^n\}^2 + 2\{X^m,K\}\{X^m,A^K\} + m^2\{X^m,H\}^2   \nonumber \\ &+  \{X^m,K\}^2  + \{X^m,A^K\}^2 +\{X^m,A^H\}^2  + m^2\{H,A^K\}^2\nonumber \\ & +  2m\{X^m,H\}\{X^m,A^H\} + 2m\{H,A^K\}\{A^H,A^K\} +  \{K,A^K\}^2  \nonumber \\ &+ 2\{A^H,K\}\{A^H,A^K\} + \{A^H,A^K\}^2 + \{K,A^H\}^2  + \{H,A^H\}^2 \bigg)\nonumber \\ &   -2P_0^+ T_{M2}  (\bar{\Psi}\Gamma^-\Gamma_m \{X^m,\Psi\} + \bar{\Psi}\Gamma^-\Gamma_K \{A^K,\Psi\}+\bar{\Psi}\Gamma^-\Gamma_H \{A^H,\Psi\} \nonumber \\ &+  \bar{\Psi}\Gamma^-\Gamma_K\{K,\Psi\} +  \bar{\Psi}\Gamma^-\Gamma_H\{H,\Psi\}) \Bigg].    
\end{align}

\begin{figure*}
    \centering
    \includegraphics[scale=0.3]{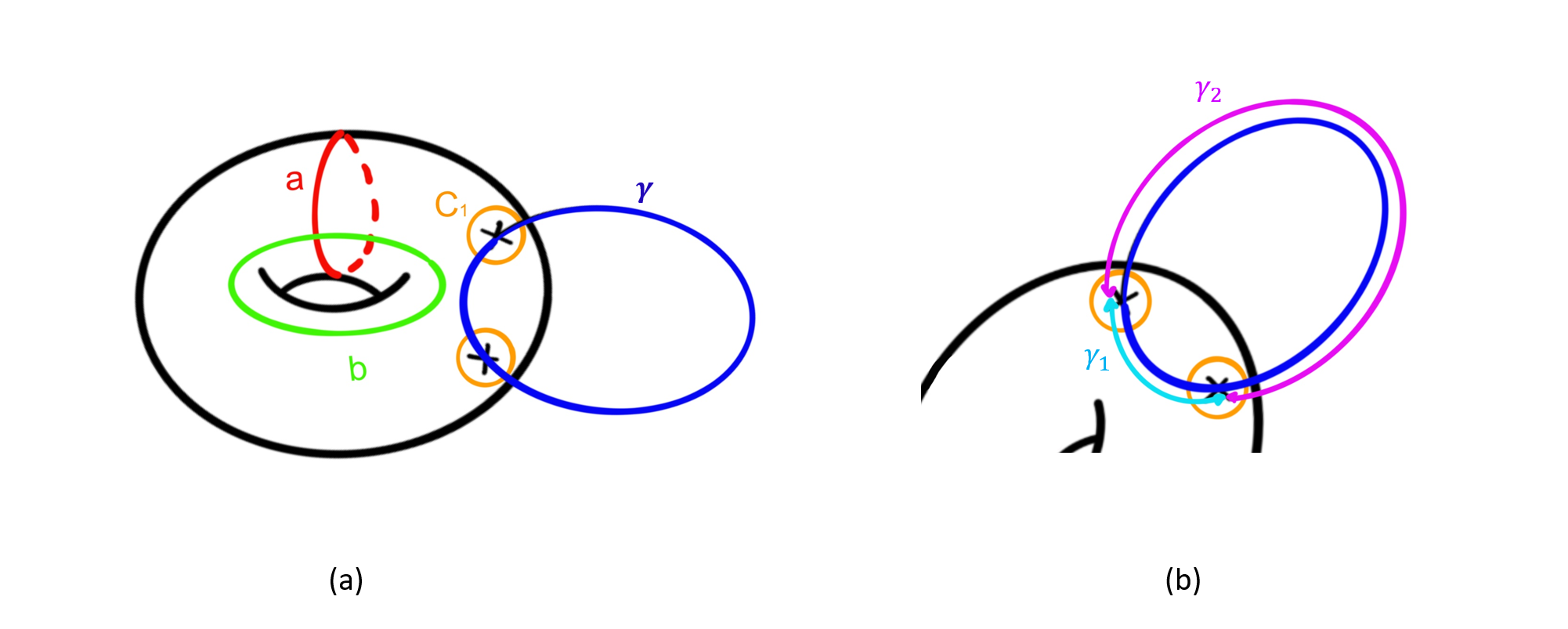}
    \caption{(a) Nontrivial cycles over $\tilde{\Sigma}_{1,2}$. (b) The curve $\gamma$ and his decomposition into the curves $\gamma_1$ and $\gamma_2$.}
    \label{fig:curves}
\end{figure*}

By defining
$$f\equiv \bigg(\frac{P_K}{\sqrt{W}}dX^K+\frac{P_H}{\sqrt{W}}dX^H+\frac{P_m}{\sqrt{W}} dX^m + \bar{\Psi}\Gamma^-d\Psi\bigg)$$, 
let us now discuss the constraints after deforming $\Sigma_2$. First, we have the local APD constraint given by
\begin{equation}
%\begin{eqnarray*}
df= 0.
%\end{eqnarray*}
\end{equation}
On the other hand, we have also four global constraints, the first two are associated with the homology basis of cycles defined over the regular torus (see figure (\ref{fig:curves}-(a))), i.e  
\begin{equation}
\zeta_1 \equiv \int_a f = 0, \qquad \mbox{and}, \qquad  
\zeta_2 \equiv \int_b f = 0
\end{equation}
We have another constraint associated to the singularities
\begin{equation}
%\begin{eqnarray*}
\zeta_3 \equiv \int_{C_1} f = 0.
%\end{eqnarray*}
\end{equation}
This constraint arises from the homology curve of $\Sigma_2$ around the handle, whose radius was sent to zero to get the string-like configuration. The final constraint is the one associated with the homology curve along the deformed handle of $\Sigma_2$, shown in figure 4, which is still present after deforming $\Sigma_2$ into $\tilde{\Sigma}_{1,2}$.

\begin{equation}\label{zeta4}
\zeta_4 \equiv \int_y f = 0.
\end{equation}
Notice, that we could not write $\zeta_4$ directly in terms of $K,H$. This is because the curve $\gamma$ is defined in both, $\Sigma_{1,2}$ and in the string attached in the punctures. Thus, it is convenient to separate the curve $\gamma$ into two pieces ((see figure (\ref{fig:curves}-(b))) and we will denote as $\gamma_1$ and $\gamma_2$. The curve $\gamma_1$ is the part of $\gamma$ defined over $\Sigma_{1,2}$ and $\gamma_2$ corresponds to the string with end points at the punctures. Now, because of (\ref{cts}), we can write
\begin{equation}
\zeta_4 =  \int_{\gamma_1} f 
=0.
\end{equation}

In the following, we will list some of the features of the massive supermembrane Hamiltonian. From Eq. (\ref{MM}) it can be seen that it is very different from a standard compactification of the M2-brane on a $S^1$. Firstly, it contains a mass term associated with the nontrivial topology of the $LCD$ on the target space given by

\bea \label{fc}
\lim_{\epsilon\to 0} \int_{\Sigma^{'}} dK\wedge d\hat{H} \frac{\alpha \ m^2}{4}\{K,\hat{H}\}^2=2\pi\alpha l \ m^2
\eea
This term can be interpreted as the uplift to ten non compact dimensions of the central charge condition proposed in \cite{Restuccia3}. In second place, it possesses non vanishing mass terms associated with the dynamics fields $X^m$,$A^K$ and $A^H$, these are

\begin{align*}
& (\partial_KX^m)^2+(\ \partial_{\hat{H}} X^m)^2 \not=0, \quad  (\partial_KA^K)^2+( \partial_{\hat{H}} A^K)^2 \not=0, 
\end{align*}
\begin{eqnarray*}
    (\partial_KA^H)^2+(\partial_{\hat{H}} A^H)^2 \not=0.
\end{eqnarray*}
Thus, the fermionic potential is dominated by the bosonic potential due to these non-vanishing quadratic contributions to the Hamiltonian. This fact, together with the structure of the rest of the potential, ensures that the Hamiltonian satisfies the discreteness sufficient condition found in \cite{mpgm12}, as formerly shown in \cite{mpgm14}.

Finally, we would like to mention that taking as a starting point a compact Riemann surface of genus two, is the simplest case, but it is not the only possibility to find massive terms in the Hamiltonian of the theory.

\section{Supersymmetric transformations}

In this section, we will analyze the supersymmetry of our formulation of the massive supermembrane. Thus, we shall follow the same procedure presented in the previous section, that is, we will begin with the M2-brane over a regular compact genus two Riemann surface. In general, the supermembrane action (in the light cone gauge) is invariant under the following supersymmetric transformations originally found in \cite{dwhn},

\begin{align}
&  \delta \tilde{X}^{M} = -2 \bar{\eta} \Gamma^M \Psi, \label{st1} \\ 
&  \delta \Psi =  \frac{1}{2} \Gamma^+ (D_0 \tilde{X}^{M} \Gamma_M +\Gamma^-) \eta + \frac{T_{M2}}{4P_0^+}\{\tilde{X}^{M},\tilde{X}^N\}\Gamma^+\Gamma_{MN} \eta \label{st2} \\
&  \delta \omega = -2\frac{T_{M2}}{P_0^+}\bar{\eta} \Psi \label{st3}, 
\end{align}
provided the following boundary terms are equal to zero 
\begin{align}\label{bt}
&\frac{P_0^+\delta L}{T_{M2}} = - \int_\mathbb{R} dt \int_{\Sigma} d\bigg[\bar{\Psi}\Gamma^-\Gamma_Md\tilde{X}^{M} \delta\Psi +2\bar{\Psi} \Gamma_{M}\Gamma^-\eta \nonumber \\ &+ 2 \bar{\Psi}\Gamma_{MN}\eta \partial_t \tilde{X}^{M} dX^N -\frac{2}{3}(\bar{\Psi}\Gamma^-d\Psi \bar{\eta}\Psi -  \bar{\eta}\Gamma^M\Psi \bar{\Psi}\Gamma^-\Gamma_M d\Psi)\bigg] \nonumber \\ & +  \lim_{\epsilon \rightarrow 0} \int_{\Sigma'} d^2\sigma \int_\mathbb{R} dt  \partial_t \bigg[ \sqrt{W}\bar{\Psi}\Gamma^-\partial \Psi -2\sqrt{W}\bar{\Psi}\Gamma^- \eta \nonumber \\ &+ \sqrt{W} \bar{\Psi}\Gamma_{MN}\eta \{\tilde{X}^{M},X^N\} \bigg] = 0,  
\end{align}
where $\eta$ is a constant spinor. 

Notice that, in this surface term, only the derivatives of the maps $X$ are displayed, which are single-valued. Thus, since $\Psi$ is also a single-valued and we are considering a compact regular Riemann surface as a base manifold, this surface term is identical to zero. Moreover, this allows us to conclude that, at least from this surface term, there are no restrictions to the supersymmetric parameter, $\eta$, when we take the limit $\Sigma_2\rightarrow \tilde{\Sigma}_{1,2}$.

On the other hand, in \cite{mpgm14}, it was shown that in order to preserve the topological term given in equation (\ref{fc}) and the mass terms in the Hamiltonian that lead to the good spectral properties of the Hamiltonian, we need to impose the following condition

\begin{eqnarray}
   \Gamma^+\left(\Gamma^- + \frac{1}{2}\Gamma_{KH}\right)\eta = 0,
\end{eqnarray}
which implies that half of the supersymmetry is broken, in distinction with the case of a supermembrane on a torus.
%%%%%%%%%%%%%%%%%%%%%%%%%%%%%%%%%%%%%%%%%%%%%%%%%%%%%%%%%%%%%%%%%%%%%%%%%%%%%%%%%%%%%%%
%%%%%%%%%%%%%%%%%%%%%%%%%%%%%%%%%%%%%%%%
\section{Supersymmetric algebra}

Following our analysis of the supersymmetric properties of the massive supermembrane, in this section we shall present the supersymmetric algebra of the massive supermembrane. Specifically, we will compute the supersymmetric charges and their Dirac brackets. As before, we will begin with the formation of the M2-brane over $\Sigma_2$. From (\ref{g2H}) we can derive the supercharge density associated with the transformations (\ref{st1}-\ref{st3})

\begin{align}
&J^0 = P_0^+ \sqrt{\tilde{W}}\bigg[2(\partial_0 \tilde{X}_M\Gamma^M+\Gamma^-) +  \frac{T_{M2}}{P_0^+}\{\tilde{X}^M,\tilde{X}^N\}\Gamma_{MN}\bigg]\Psi.    
\end{align}
Thus the supersymmetric charges, defined as

\begin{eqnarray}
Q^{\pm} = \frac{1}{2}\Gamma^{\pm}\Gamma^{\mp}Q, \quad Q=\int_{\Sigma_2}d\sigma^2 J^0,
\end{eqnarray}
can be written as

\begin{align}
&Q^+ = \int_{\Sigma_2} d\sigma^2 [2\tilde{P}_M\Gamma^M + T_{M2}\sqrt{\tilde{W}}\{\tilde{X}^{M},\tilde{X}^N\}\Gamma_{MN}]\tilde{\Psi}, \\
&Q^- = 2P_0^+ \Gamma^- \int_{\Sigma_2} d\sigma^2 \sqrt{\tilde{W}}\tilde{\Psi}.  
\end{align}
The only non trivial Dirac's brackets in our case (arising from the standard Dirac approach in the presence of second class constraints), are given by
\begin{align}
 &\{\tilde{X}^{M}(\sigma),P_N(\sigma')\}_{D.B} = \delta^M_N \delta^2(\sigma - \sigma'), \\
&\{\Psi^\alpha(\sigma),\Psi_\beta(\sigma')\}_{D.B} = \frac{1}{4\sqrt{\tilde{W}}P_0^+}(\Gamma^+)^\alpha_\beta \delta^2(\sigma -\sigma'), 
\end{align}
where we are considering that $\sigma$ and $\sigma'$ are the coordinates of two points inside $\Sigma_2$. With these expressions and using the Gamma matrices properties (see \cite{vp}) we get

\begin{align}
 &\{Q^-_\alpha,Q^-_\beta\}_{D.B} = -2 P_0^+ (\Gamma^+)^{\alpha \beta}, \label{db1} \\
&\{Q^+_\alpha,Q^-_\beta\}_{D.B} = -(\Gamma_M\Gamma^+\Gamma^-)_{\alpha \beta} P^M_0 \nonumber \\ &  \qquad \quad - \frac{T_{M2}}{2}(\Gamma_{MN}\Gamma^+\Gamma^-)_{\alpha \beta} \int_{\Sigma_2}d\sigma^2 \sqrt{\tilde{W}}\{\tilde{X}^{M},\tilde{X}^N\}, \label{db2} \\
&\{Q^+_\alpha,Q^+_\beta\}_{D.B} = 2(\Gamma^+)_{\alpha \beta}H - 2T_{M2}(\Gamma^+\Gamma_M)_{\alpha \beta}  \int_{\Sigma_2}d\sigma^2 \sqrt{\tilde{W}}\{\tilde{X}^-,\tilde{X}^{M}\}. \label{db3}   
\end{align}
Notice that this is the most general form of the supersymmetric algebra for the supermembrane found in \cite{deWit2,deWit3}. Now we can analyze the surface terms in detail. However, since we are considering the limit $\Sigma_2 \rightarrow \tilde{\Sigma}_{1,2}$, the surface term in the last two terms leads to several differences. This is due to the two singular points resulting from the deformation of $\Sigma_2$. From the general superalgebra in eleven dimensions (see for example \cite{vanProeyen,Townsend4}) it can be seem that the surface terms can be interpreted in terms of tensorial charges. Specifically, the surface term in (\ref{db2}) and (\ref{db3}) are related to the charges $Z_{MN}$ and $Z_{+M}$, respectively. As is discussed in \cite{Hull10}, the 2-form $Z_{MN}$ gives a 2-brane charge, and it has been conjectured that the dual of the from $Z_{+M}$ gives a 9-brane charge.  

Now, let us analyze in detail the surface terms beginning with the one in (\ref{db2}). In the limit $\Sigma_2 \rightarrow \tilde{\Sigma}_{1,2}$ it can be shown that 

\begin{align}
     &\int_{\Sigma_2}d\sigma^2 \sqrt{\tilde{W}}\{\tilde{X}^{M},\tilde{X}^N\}  \rightarrow  \int_{\tilde{\Sigma}_{1,2}}d\sigma^2 \sqrt{W}\{X^{M},X^N\}. 
\end{align}
Thus, following the same arguments of section $IV$, we can also write 
\begin{align*}
&\int_{\tilde{\Sigma}_{1,2}}d\sigma^2 \sqrt{W}\{X^M,X^N\} = \int_{\Sigma_{1,2}}d\sigma^2 \sqrt{W}\{X^M,X^N\} \\ 
& = \lim_{\epsilon \rightarrow 0} \int_{\Sigma'}d\sigma^2 \sqrt{W}\{X^M,X^N\}.    
\end{align*}
The only non-trivial contributions of this term are given by 
\begin{align*}
&\lim_{\epsilon \rightarrow 0} \int_{\Sigma'}d\sigma^2 \sqrt{W}\{X^M,X^N\}  = \frac{l\alpha}{4\pi} \lim_{\epsilon \rightarrow 0} \int_{\Sigma'}[\delta^M_K (\frac{1}{2}d\hat{K}\wedge d\hat{H}  \nonumber \\ & + d\hat{A^K}\wedge  d\hat{H}) + \delta_m^M d\hat{X^m}\wedge  d\hat{H}]\delta^N_H - (M\rightarrow N).    \end{align*}

This can be simplified to obtain 
\begin{align}\label{bt2}
&\lim_{\epsilon \rightarrow 0} \int_{\Sigma'}d\sigma^2 \sqrt{W}\{X^M,X^N\} \nonumber \\ & = l\alpha \bigg[\frac{1}{2}\delta^M_K + \bigg(\delta_m^M \int_{\gamma_1}dX^m + \delta_K^M \int_{\gamma_1}dA^K\bigg)\bigg]\delta^N_H  -(M\rightarrow N).    
\end{align}

Now, we can a analyze the surface term in (\ref{db3}). Following the same idea of the previous case, we can write (in the limit $\Sigma_2 \rightarrow \tilde{\Sigma}_{1,2}$)

\begin{eqnarray*}
    \int_{\tilde{\Sigma}_{1,2}}d\sigma^2 \sqrt{W}\{X^-,X^M\} = \int_{\Sigma_{1,2}}d\sigma^2 \sqrt{W}\{X^-,X^M\},
\end{eqnarray*}
which leads to

\begin{align}\label{sf23}
&\int_{\Sigma'}d\sigma^2 \sqrt{W}\{X^-,X^M\} =  \bigg(\lim_{\epsilon \rightarrow 0} \int_{\Sigma'} X^M \phi + \zeta_2\int_a dX^M - \zeta_1\int_b dX^M  \nonumber  \\ &  + \frac{l\alpha}{4\pi}\lim_{\epsilon \rightarrow 0} \bigg[\sum_r\bigg(\int_{C_r}+\int_{D_r}\bigg) + \sum_u\bigg(\int_{I_u} + \int_{I_{u^{-1}}}\bigg)\bigg]X^MdX^- \bigg).    
\end{align}
Since $X^MdX^-$ is well defined at $P_r$, the limit $\epsilon \rightarrow 0$ of the integral over $D_r$ is equal to zero. The integrals around the punctures lead to 

\begin{eqnarray*}
\lim_{\epsilon \rightarrow 0}\sum_r\int_{C_r} X^MdX^- = -\int_{\gamma_1}dX^M \int_{C_1} dX^- = -\int_{\gamma_1}dX^M \zeta_3. \nonumber \\
\end{eqnarray*}
Moreover, it can be proved that

\begin{eqnarray*}
&&\sum_u\bigg(\int_{I_u} + \int_{I_{u^{-1}}}\bigg)\bigg]X^MdX^- = - 2\pi \alpha \delta^M_H \int_{\gamma_1} dX^-. \nonumber \\ &&
\end{eqnarray*}

Thus, the final form of the massive supermembrane algebra is given by 

\begin{align}
 &\{Q^-_\alpha,Q^-_\beta\}_{D.B} = -2 P_0^+ (\Gamma^+)^{\alpha \beta}, \label{db12} \\
&\{Q^+_\alpha,Q^-_\beta\}_{D.B} = -(\Gamma_M\Gamma^+\Gamma^-)_{\alpha \beta} P^M_0  - \frac{T\alpha l}{2}(\Gamma_{MH}\Gamma^+\Gamma^-)_{\alpha \beta}\delta^M_K, \label{db22} \\
&\{Q^+_\alpha,Q^+_\beta\}_{D.B} = 2(\Gamma^+)_{\alpha \beta}H - 2T(\Gamma^+\Gamma_M)_{\alpha \beta} \bigg[\lim_{\epsilon \rightarrow 0} \int_{\Sigma'} X^M \phi_1 \nonumber \\ & + \frac{2\pi \alpha \delta^M_H}{Im(\tau)} \bigg(Im(Z_2-Z_1)\zeta_2 - Im((Z_2-Z_1)\bar{\tau})\zeta_1 \bigg) \nonumber \\ &  - 2 \bigg(l \delta^M_K \zeta_3 + \pi \alpha \delta^M_H \zeta_4 \bigg)  \bigg]. \label{db32}   
\end{align}
At this point, the following comments are in order:

\begin{itemize}
    \item In \cite{mpgm14}, the massive supermembrane is interpreted as the uplift to ten non compact dimensions of the supermembrane with $C_{\pm}$ fluxes and parabolic monodromy. Then, we can interpret (\ref{db12})-(\ref{db32}) as the generalization of the M2-brane super algebra when the world volume of the theory is a twice punctured torus.
    \item In (\ref{db22}), we get a constant term that is analogous to the fluxes/central charge contribution to the super algebra presented in \cite{mpgm10}. However, in the present case, this term is not proportional to an integer.
    \item We showed that the surface term in (\ref{db32}), can be written in terms of the constraints of the theory. The terms related to the constraints are analogous to the case without punctures (see \cite{mpgm10}). However, in our case, we have two extra global constraints related to the punctures. Moreover, the multiplicative factors of each are related to the moduli of the twice punctured torus, while in \cite{mpgm10} are the winding numbers of the theory. 
\end{itemize}

\section{Area preserving diffeomorphisms}

Another relevant symmetry of the supermembrane theory is the invariance under APD. In this section, we will discuss the realization of this symmetry in the massive supermembrane formulation. As discussed in previous sections, the Hamiltonian of the supermembrane on $\tilde{\Sigma}_{1,2}$ is the same as in $\Sigma_{1,2}$. Thus we will restrict ourselves to the analysis of the APD for the Hamiltonian given by (\ref{MM}) . Under APD connected to the identity,  any functional  O of the canonical variables transforms as

\begin{eqnarray}
\delta_\xi O = \left\lbrace O, <d\xi \wedge \left(\frac{P_M}{\sqrt{W}}dX^M + \bar{\Psi}\Gamma^-d\Psi\right)> \right\rbrace_{P.B.},
\end{eqnarray}
where, in this case corresponds to 
\begin{align}
 & <d\xi \wedge \left(\frac{P_M}{\sqrt{W}}dX^M + \bar{\Psi}\Gamma^-d\Psi\right)> = \lim_{\epsilon \rightarrow 0} \int_{\Sigma'}d\xi \wedge \left(\frac{P_M}{\sqrt{W}}dX^M + \bar{\Psi}\Gamma^-d\Psi\right).   
\end{align}
In these expressions,  $\xi$, is the infinitesimal parameter of the transformation. This parameter defines globally a closed 1-form $d\xi$. Thus, $\xi$ is globally defined over $\Sigma'$, that is, the $d\xi$ is an exact form. If $\xi$ is not globally defined, then $d\xi$ is a closed but not exact form.  It can be verified that the  following APD transformations hold for the massive supermembrane
\begin{align}
  \delta_\xi X^M = \{\xi,X^M\}, \quad \delta_\xi P_M = \sqrt{W} \left\lbrace \xi, \frac{P_M} {\sqrt{W}}\right\rbrace, \quad \delta_\xi \Psi = \{\xi,\Psi\}.  
\end{align}
They are the same as the ones found in \cite{dwhn} and \cite{deWit4} describing the case of the supermembrane on a flat Minkowski spacetime. They also hold for the supermembrane with central charge \cite{mpgm}.
Now, in order to determine the symmetries of the massive supermembrane under area preserving diffeomorphims non connected to the identity, we shall start by recalling the non punctured case. For these transformations, the homology basis defined over a two torus without punctures transforms as

\begin{eqnarray}
d\hat{X}^i \rightarrow  S^i_j d\hat{X}^j, \quad S \in Sl(2,\mathbb{Z}),
\end{eqnarray}
while
\begin{eqnarray}
\tau \rightarrow \frac{a\tau + b}{c\tau +d}, \quad \left(\begin{array}{cc}
    a & b \\
    c & d
\end{array}\right) \in Sl(2,\mathbb{Z}).
\end{eqnarray}
Thus, from (\ref{Mm}), it can be found that, under these transformations, the Mandelstam map transforms as 
\begin{eqnarray}
&& F\left(\frac{z}{c\tau+d},\frac{Z_1}{c\tau+d},\frac{Z_2}{c\tau+d},\frac{a\tau + b}{c\tau +d}\right) \nonumber \\ && \qquad \qquad = F(z,Z_1,Z_2,\tau) + \frac{i\pi c }{c\tau +d} (Z_1^1-Z_2^2),
\end{eqnarray}
implying that $dF$ (and therefore $dG$ and $dH$) is invariant under APD connected and non connected to the identity. However, the 1-form $dK$ is invariant under the APD connected to the identity but it may not be invariant under the not connected to the identity transformations. Indeed, we get 

\begin{align}
dK \rightarrow \frac{1-K_0^2}{(1+K_0 K)^2}dK, \quad K_0 = tanh\left[Re\left(\frac{i\pi c }{c\tau +d} (Z_1^1-Z_2^2)\right)\right].    
\end{align}
It is clear that the massive supermembrane action will be invariant under APD non connected to the identity as long as $dK$ is invariant under these transformations. Thus, the only possible transformation in $Sl(2,\mathbb{Z})$ that satisfy this requirement is when $c=0$. In other words, the massive supermembrane is invariant under APD connected to the identity, but it is only invariant under the parabolic subgroup of $Sl(2,\mathbb{Z})$, transforming isotopy classes of not connected to the identity APD. The massive supermembrane discussed in this work (see also  \cite{mpgm14}), represents an explicit realization of Hull's conjecture about the origin, in M-theory, of Roman's supergravity in terms of torus bundles with parabolic $Sl(2,\mathbb{Z})$ monodromy. 

In \cite{mpgm14}, it was presented the relation between the monodromies defined over a twice punctured torus and the nontrivial (1,1)-Knots. This relation is based on an epimorphism $\Omega$ between the mapping class group of the twice punctured torus ($MCG(\Sigma_{1,2})$)  and the mapping class group of the regular torus ($MCG(\Sigma) $)
    
    \begin{equation}
    \Omega: MCG(\Sigma_{1,2}) \rightarrow MCG(\Sigma) \cong Sl(2,\mathbb{Z}).
    \end{equation}
    
Now, since our M2-brane formulation is only invariant under the $Sl(2,\mathbb{Z})$ parabolic subgroup, the monodromies are also restricted to this subgroup as shown in \cite{mpgm14}. Furthermore, this could be classified by all the non trivial (1,1)-Knots that, under $\Omega$, are mapped into the parabolic subgroup of $Sl(2,\mathbb{Z})$.

\section{Conclusions}

We obtained the supersymmetric algebra of the massive Supermembrane with target space $M_9X LCD$ and base manifold a punctured torus. The $LCD$ is taken to be conformally equivalent to a punctured torus. The target space has ten non-compactified dimensions and a nontrivial compactification on the 11th one. The compactified dimension is not homeomorphic  to a circle. The worldvolume considered corresponds to a 2-genus Riemann surface where a zero limit radius has been imposed on one homology cycle. The Hamiltonian of this construction shows in an explicit way the role of surface terms generated by the singularities. The surface terms are expressed in terms of the local and four global APD constraints. The construction can be generalized to more punctures, although the explicit construction will become more cumbersome. We also discuss the invariance of the massive M2-brane under APD. We also show, using a different argument than the one in \cite{mpgm14}, that only parabolic $Sl(2,\mathbb{Z})$ symmetry among isotopy classes is preserved , in agreement with Hull's conjecture about the M-theory origin of 10D massive Romans supergravity.

\section{Declaration of competing interest}
The authors declare that they have no known competing financial interests or personal relationships that could have appeared to influence the work reported in this paper.

\section{Data availability}
No data was used for the research described in the article.

\section{Acknowledgements}

P.L. has been supported by the projects MINEDUC-UA ANT1956, MINEDUC-UA ANT2156 of the U. de Antofagasta. P.L and A.R have been supported by the MINEDUC-UA ANT2255 of the U. de Antofagasta. The authors also thank to Semillero funding project SEM18-02 from U. Antofagasta.


\begin{thebibliography}{00}
\bibitem{dwln}
B.~De Wit, M.~Lüscher, and H.~Nicolai.
\newblock The supermembrane is unstable.
\newblock {\em Nuclear Physics B}, 320(1):135 -- 159, 1989.

\bibitem{Nicolai2}
Olaf Lechtenfeld and Hermann Nicolai.
\newblock {A perturbative expansion scheme for supermembrane and matrix
  theory}.
\newblock {\em JHEP}, 02:114, 2022.

\bibitem{mpgm15}
L.~Boulton, M.P. Garcia~del Moral, and A.~Restuccia.
\newblock {Existence of a supersymmetric massless ground state of the $SU(N)$
  matrix model globally on its valleys}.
\newblock {\em JHEP}, 05:281, 2021.

\bibitem{mpgm}
M.~P. Garcia~del Moral and A.~Restuccia.
\newblock {Spectrum of a noncommutative formulation of the D = 11 supermembrane
  with winding}.
\newblock {\em Phys. Rev.}, D66:045023, 2002.

\bibitem{Dasgupta}
K.~Dasgupta, M.~M. Sheikh-Jabbari, and M.~Van~Raamsdonk.
\newblock {Matrix perturbation theory for M theory on a PP wave}.
\newblock {\em JHEP}, 05:056, 2002.

\bibitem{mpgm6}
M.~P. Garcia Del~Moral, C.~Las~Heras, P.~Leon, J.~M. Pena, and A.~Restuccia.
\newblock {M2-branes on a constant flux background}.
\newblock {\em Phys. Lett.}, B797:134924, 2019.

\bibitem{mpgm14}
M.P. Garcia~del Moral, P.~Leon, and A.~Restuccia.
\newblock {The massive supermembrane on a knot}.
\newblock {\em JHEP}, 10:212, 2021.

\bibitem{Maldacena3}
D.~E. Berenstein, J.~M. Maldacena, and H.~S. Nastase.
\newblock {Strings in flat space and pp waves from N=4 superYang-Mills}.
\newblock {\em JHEP}, 04:013, 2002.

\bibitem{duff}
M.~J.~Duff, P.~S.~Howe, T.~Inami and K.~S.~Stelle,
\newblock {Superstrings in D=10 from Supermembranes in D=11}
\newblock {\em Phys. Lett. B}, 191,70 , 1987.

\bibitem{lc}
M.~P.~G.~del Moral, C.~las Heras and A.~Restuccia,
\newblock {Type IIB parabolic ($p,q$)-strings from M2-branes with fluxes}
%\newblock {\em Phys. Lett. B}, 191,70 , 1987.

\bibitem{Bergshoeff7}
E. Bergshoeff and J.~P. van~der Schaar.
\newblock {On M nine-branes}.
\newblock {\em Class. Quant. Grav.}, 16:23--39, 1999.

\bibitem{dwhn}
B.~de~Wit, J.~Hoppe, and H.~Nicolai.
\newblock On the quantum mechanics of supermembranes.
\newblock {\em Nucl. Phys. B}, 305(4):545 -- 581, 1988.

\bibitem{Bergshoeff11}
E.~Bergshoeff, J.~Gomis, B.~Rollier, J.~Rosseel and T.~ter Veldhuis,
\newblock{Carroll versus Galilei Gravity}
\newblock{\em JHEP} 03:165, 2017

\bibitem{Gomis20}
J.~Gomis, A.~Kleinschmidt, J.~Palmkvist and P.~Salgado-Rebolledo,
\newblock{Symmetries of post-Galilean expansions}
\newblock{\em Phys. Rev. Lett.} 124, 8, 081602, 2020.

\bibitem{Bergshoeff17}
R.~Andringa, E.~Bergshoeff, S.~Panda and M.~de Roo,
\newblock{Newtonian Gravity and the Bargmann Algebra}
\newblock{\em Class. Quant. Grav.} 28:105011, 2011.

\bibitem{M-algebra}
M.~Hassaine, R.~Troncoso and J.~Zanelli,
\newblock{11D supergravity as a gauge theory for the M-algebra}
\newblock{\em PoS \textbf{WC2004}} 006, 2005.

\bibitem{Edelstein06}
J.~D.~Edelstein, M.~Hassaine, R.~Troncoso and J.~Zanelli,
\newblock{Lie-algebra expansions, Chern-Simons theories and the Einstein-Hilbert Lagrangian}
\newblock{\em Phys. Lett. B} 640:78-284, 2006.

\bibitem{Ravera22}
L.~Ravera and U.~Zorba,
\newblock{Carrollian and Non-relativistic Jackiw-Teitelboim Supergravity}
\newblock{} 2022.


\bibitem{Bergshoeff}
E.~Bergshoeff, E.~Sezgin, and P.K. Townsend.
\newblock Supermembranes and eleven-dimensional supergravity.
\newblock {\em Phys. Lett. B}, 189(1):75-78, 1987.

\bibitem{Mandelstam}
S.~Mandelstam.
\newblock Interacting-string picture of dual-resonance models.
\newblock {\em Nuclear Physics B}, 64:205 -- 235, 1973.

\bibitem{Giddings2}
S.~B. Giddings and S.~A. Wolpert.
\newblock A triangulation of moduli space from light-cone string theory.
\newblock {\em Comm. Math. Phys.}, 109(2):177--190, 1987.

\bibitem{Nicolai}
H. Nicolai and R. Helling.
\newblock {Supermembranes and M(atrix) theory}.
\newblock In {\em {Nonperturbative aspects of strings, branes and
  supersymmetry. Proceedings, Spring School on nonperturbative aspects of
  string theory and supersymmetric gauge theories and Conference on
  super-five-branes and physics in 5 + 1 dimensions, Trieste, Italy, March
  23-April 3, 1998}}, pages 29--74, 1998.

\bibitem{farkas}
H.M. Farkas and I.~Kra.
\newblock {\em Riemann Surfaces}.
\newblock Graduate Texts in Mathematics. Springer New York, 2012.


%\cite{Martin:1997cb}
\bibitem{Restuccia3}
I.~Martin, A.~Restuccia and R.~S.~Torrealba,
\newblock {On the stability of compactified D = 11 supermembranes}.
\newblock {\em Nucl. Phys. B}, 521, 117-128, 1998.
%41 citations counted in INSPIRE as of 01 Jan 2023

\bibitem{mpgm12}
L. Boulton, M.P. Garcia~del Moral, and Alvaro Restuccia.
\newblock {Spectral properties in supersymmetric matrix models}.
\newblock {\em Nucl. Phys. B}, 856:716--747, 2012.

%\cite{VanProeyen:1999ni}
\bibitem{vp}
A.~Van Proeyen.
\newblock {Tools for supersymmetry}
\newblock {\em Ann. U. Craiova Phys.}, 9, 1-48, 1999.

\bibitem{deWit2}
B.~de~Wit, J.~Hoppe, and H.~Nicolai.
\newblock {On the Quantum Mechanics of Supermembranes}.
\newblock {\em Nucl. Phys.}, B305:545, 1988.
\newblock [,73(1988)].

\bibitem{deWit3}
Bernard de~Wit, Kasper Peeters, and Jan~C. Plefka.
\newblock {Open and closed supermembranes with winding}.
\newblock {\em Nucl. Phys. B Proc. Suppl.}, 68:206--215, 1998.

\bibitem{vanProeyen}
J~W van Holten and A~van Proeyen.
\newblock N=1 supersymmetry algebras in d=2,3,4 mod 8.
\newblock {\em Journal of Physics A: Mathematical and General},
  15(12):3763--3783, dec 1982.

\bibitem{Townsend4}
P.~K. Townsend.
\newblock {P-brane democracy}.
\newblock In {\em {PASCOS / HOPKINS 1995 (Joint Meeting of the International
  Symposium on Particles, Strings and Cosmology and the 19th Johns Hopkins
  Workshop on Current Problems in Particle Theory)}}, pages 375--389, 7 1995.

\bibitem{Hull10}
C.~M. Hull.
\newblock {Gravitational duality, branes and charges}.
\newblock {\em Nucl. Phys. B}, 509:216--251, 1998.

\bibitem{mpgm10}
M.P. Garcia~del Moral, C.~Las~Heras, P.~Leon, J.M. Pena, and A.~Restuccia.
\newblock {Fluxes, twisted tori, monodromy and $U(1)$ supermembranes}.
\newblock {\em JHEP}, 09:097, 2020.

\bibitem{deWit4}
B.~de~Wit, U.~Marquard, and H.~Nicolai.
\newblock {Area Preserving Diffeomorphisms and Supermembrane Lorentz
  Invariance}.
\newblock {\em Commun. Math. Phys.}, 128:39, 1990.

\end{thebibliography}
\end{document}